\documentclass[12pt]{iopart}
\usepackage{iopams,amstext,bm,braket}
\usepackage[breaklinks=true,colorlinks=true,linkcolor=blue,urlcolor=blue,citecolor=blue]{hyperref}
\usepackage{graphicx}
\usepackage{dcolumn}
\usepackage{verbatim}
\usepackage{subfigure}
\usepackage{color}
\usepackage{float}

\newcommand{\sinput}{s_\text{in}(t)}
\newcommand{\sind}{s_\text{in}^\dagger(t)}

\newcommand{\snout}{s_\text{out}(t)}

\newcommand{\bin}{b_\text{in}(t)}
\newcommand{\bout}{b_\text{out}(t)}
\newcommand{\bind}{b_\text{in}^\dagger(t)}
\newcommand{\base}{b_\text{ASE}(t)}
\newcommand{\brase}{b_\text{RASE}(t)}
\newcommand{\din}{d_\text{in}(t)}
\newcommand{\dout}{d_\text{out}(t)}
\newcommand{\gaone}{\gamma_\text{a,1}}
\newcommand{\gbone}{\gamma_\text{b,1}}
\newcommand{\gatwo}{\gamma_\text{a,2}}
\newcommand{\gbtwo}{\gamma_\text{b,2}}
\newcommand{\xa}{x_\text{A}}
\newcommand{\xb}{x_\text{B}}
\newcommand{\pa}{p_\text{A}}
\newcommand{\pb}{p_\text{B}}

\begin{document}
\title{Cavity enhanced rephased amplified spontaneous emission}

\author{Lewis A Williamson and Jevon J Longdell}
\address{The Jack Dodd Centre for Quantum Technology, Department of Physics, University of Otago, Dunedin, New Zealand.}
\ead{jevon.longdell@otago.ac.nz}

\begin{abstract}
Amplified spontaneous emission is usually treated as an incoherent noise process. Recent theoretical and experimental work using rephasing optical pulses has shown that rephased amplified spontaneous emission (RASE) is a potential source of wide bandwidth time-delayed entanglement. Due to poor echo efficiency the plain RASE protocol doesn't in theory achieve perfect entanglement. Experiments done to date show a  very small amount of entanglement at best. Here we show that rephased amplified spontaneous emission can, in principle, produce perfect multimode time-delayed two mode squeezing when the active medium is placed inside a Q-switched cavity. 
\end{abstract}

\maketitle

\section{Introduction}
The bitrate available from current quantum networks falls of very quickly with increasing attenuation in the transmission path. Quantum repeaters~\cite{briegel1998quantum} stand to alleviate this problem. The proposal of Duan-Lukin-Cirac and Zoller (DLCZ)~\cite{dlcz} first suggested the use of atomic ensembles to generate time-separated entangled photons. There has been a lot of experimental progress in the use of atomic ensembles~\cite{matsukevich2005entanglement,felinto2006conditional,chaneliere2007quantum,yuan2007synchronized,laurat2007heralded,
yuan2008experimental,choi2008mapping,clausen2011quantum}, but not yet a practical quantum repeater.

\begin{figure}[h]
\begin{center}
\includegraphics[width=\textwidth]{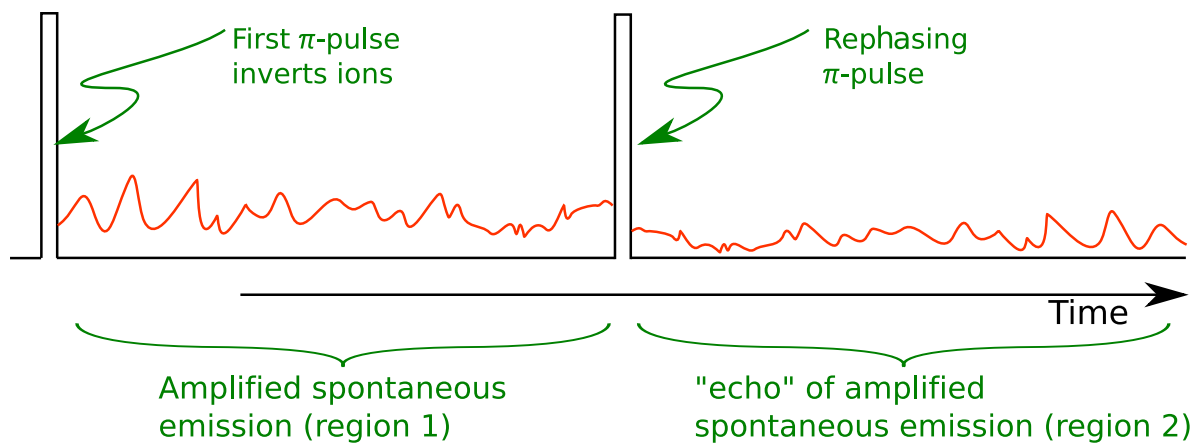}
\end{center}
\caption{\label{RASEscheme}The RASE scheme. The first $\pi$-pulse shifts the two-level atoms into their excited state. Spontaneous emission from the excited atom ensemble produces the amplified spontaneous emission (ASE). During this emission, the atoms continually dephase due to inhomogeneous broadening. The rephasing $\pi$-pulse inverts the excitation of and rephases the atoms. Atoms that contributed to the ASE continue to emit photons to produce the RASE. This results in RASE that is a time-reversed ``echo'' of the ASE. The RASE and ASE are correlated at times equally spaced from the rephasing $\pi$-pulse.}
\end{figure}

Quantum memories, like the DLCZ protocol, use the large interactions possible between an optical field and a collective excitation in an ensemble. Early work on quantum memories proposed using photon echo techniques and inhomogeneously broadened atoms to efficiently store and retrieve quantum states of light~\cite{moiseev2001complete,moiseev2003quantum,nilsson2005solid,kraus2006quantum} (for a review, see~\cite{afzelius2010photon}). Since these initial proposals, there has been impressive demonstrations of quantum memories using photon echo techniques~\cite{hosseini2009coherent,hedges2010efficient,afzelius2010demonstration,saglamyurek2011broadband,hosseini2011unconditional,
sabooni2013efficient,gundougan2013coherent,timoney2013single,bonarota2013photon,jobez2014cavity}. A distinct advantage such techniques have is that they are multimode~\cite{ simon2007quantum,moiseev2010efficient}. Ledingham et al.~\cite{ledingham2010nonclassical} showed how rephased amplified spontaneous emission (RASE) can be used as a source of photon streams with time-separated entanglement. This suggested that photon echoes were not only useful for making the quantum memories used in quantum repeaters; they could also be used as the entanglement source. The RASE is illustrated in figure~\ref{RASEscheme}.

Realization of the RASE scheme has been demonstrated in two separate systems. One system follows the original proposal~\cite{ledingham2012experimental} and the other~\cite{beavan2012demonstration} implements a modified approach based on four-level atoms~\cite{beavan2011photon}. Both demonstrations show that correlations exist between the ASE field and the RASE field. In the case of~\cite{ledingham2012experimental} the correlation was strong enough and the noise was low enough to show evidence of entanglement, although the confidence level was not high. 

As a source of entanglement, RASE as it was  initially proposed in~\cite{ledingham2010nonclassical} and~\cite{beavan2011photon} is imperfect. The main problem is the efficiency of the recall of entanglement from the atomic ensemble.
In the first step of~\cite{ledingham2010nonclassical} illustrated in figure~\ref{RASEscheme},  an inverted ensemble of atoms creates ASE. This light is entangled with collective degrees of freedom of the atomic ensemble. The excitation in those collective degrees of freedom is then recalled into an output light field with a rephasing $\pi$-pulse. The fact that the quality of entanglement was limited by this recall efficiency is problematic. The low optical depths desirable in the first step to get weak ASE lead to poor recall efficiency. The four-level RASE~\cite{beavan2011photon} is slightly better in that separate transitions are used for the ASE and RASE steps. This means that a weak transition can be used for the ASE step and a stronger one for the RASE.
Recently~\cite{stevenson2013single} further improvement has been suggested by tailoring the spatial density of the ions.

The recall efficiency of the RASE scheme can be improved by placing the atoms inside a low finesse cavity, an approach also used in quantum memories~\cite{afzelius2010impedance, moiseev2010efficient, sabooni}. In this paper we examine this  process of cavity enhanced rephased amplified spontaneous emission (CRASE). We will show that in the appropriate regime the CRASE scheme is capable of achieving a recall efficiency of 100\%, and in principle perfect multimode, time-separated entanglement.

\section{The Hamiltonian for our system}
The evolution of CRASE is qualitatively the same as the RASE evolution shown in figure~\ref{RASEscheme}. The interaction picture Hamiltonian for our system takes the form (making the rotating wave approximation and setting $\hbar\equiv 1$)
\begin{equation}
H=H_1+H_2
\end{equation}
where
\begin{equation}\label{H1}
H_1=\sum_{k=1}^N\sigma_+^k(t)\sigma_-^k(t)+\rmi g\sum_{k=1}^N(\sigma^k_+(t)a(t)-\sigma^k_-(t)a^\dagger(t))
\end{equation}
 is the usual Jaynes-Cummings Hamiltonian that models the interaction between $N$ two-level atoms and the cavity mode and
\begin{equation}\label{eq:H2}
\fl H_2=\int_{-\infty}^\infty\Delta b^\dagger(\Delta,t)b(\Delta,t)\,\rmd\Delta+\rmi\int_{-\infty}^\infty\kappa(\Delta)\left(b^\dagger(\Delta,t)a(t)-b(\Delta,t)a^\dagger(t)\right)\,\rmd\Delta
\end{equation}
models the interaction between the external radiation field and the cavity~\cite{gardiner2004quantum,dutra2005cavity}. The operators $\sigma^k_+(t)$ and $\sigma^k_-(t)$ are raising and lowering operators respectively for atom $k$, $a(t)$ is the destruction operator for the cavity mode, $g$ is the coupling between the atoms and the cavity mode, which we take to be uniform, $b(\Delta,t)$ are destruction operators for the external radiation modes, $\Delta$ is the detuning from the cavity mode frequency and $\kappa(\Delta)$ is the coupling between the radiation mode with detuning $\Delta$ and the cavity mode.

\section{The ASE field}
At the beginning of region 1 in figure~\ref{RASEscheme} the atoms are in their excited state. Spontaneous emission by the atoms produces the ASE field. We will assume that the atoms are weakly coupled to the cavity (small $g$) so that the atoms remain predominantly in their excited state throughout region 1. This allows us to approximate each atom as an \emph{inverted} harmonic oscillator by setting $\sigma_+^k(t)\rightarrow s_k(t)$, where $s_k(t)$ are destruction operators satisfying $[s_k(t),s_{k^\prime}^\dagger(t)]=\delta_{kk^\prime}$~\cite{gardiner2004quantum}. Like the ordinary harmonic oscillator, the eigenstates of an inverted harmonic oscillator form a ladder of equally spaced energy states. The ladder is inverted in the sense that an energy eigenstate $\ket{n}$  contains $n$ units of \emph{negative energy} so that the state $\ket{n+1}\propto s_k^\dagger \ket{n}$ has a lower energy than the state $\ket{n}$. We will also approximate the collection of harmonic oscillators by a continuous field by setting $\sqrt{N}s_k(t)\rightarrow s(\Delta,t)$. The operators $s(\Delta,t)$ are destruction operators for a collective excitation across oscillators with detuning $\Delta$ and satisfy $[s(\Delta,t),s^\dagger(\Delta^\prime,t)]=\delta(\Delta-\Delta^\prime)$ in the limit $N\rightarrow \infty$. The Hamiltonian $H_1$ then takes the form
\begin{equation}
\fl H_1=-\int_{-\infty}^\infty \Delta s^\dagger(\Delta,t)s(\Delta,t)\,\rmd\Delta+\rmi\int_{-\infty}^\infty g(\Delta)\left(s(\Delta,t)a(t)-s^\dagger(\Delta,t)a^\dagger(t)\right)\,\rmd\Delta\label{eq:H1b}
\end{equation}
where $\sqrt{N}g\rightarrow g(\Delta)$.

Input-output theory~\cite{gardiner2004quantum} is often used when describing quantum systems interacting with a continuum of radiation modes, such as the situation described by equation~\eref{eq:H2}. The interaction between the excited state atoms and the cavity mode,  described by equation~\eref{eq:H1b}, is very similar: in both cases the cavity is interacting with a continuum of harmonic oscillators. This allows us to use input-output theory for the atom-cavity interaction also.

Following the standard input-output treatment we assume $\kappa(\Delta)$ and $g(\Delta)$ are slowly varying for our range of frequencies of interest. We can then make the \emph{first Markov approximation} by setting $\kappa(\Delta)\rightarrow \kappa(0)\equiv \sqrt{\gbone/2\pi}$ and $g(\Delta)\rightarrow g(0)\equiv \sqrt{\gaone/2\pi}$. In the case of $g$, this is valid when the inhomogeneous broadening of the atoms is much broader than the cavity bandwidth. The loss rate of the bare cavity is $\gbone$. We call $\gaone$ the `gain rate' of the cavity; this is the rate describing the exponential growth of light in the cavity if the cavity mirrors were perfect. We will work in the regime where $\gaone<\gbone$ so that we are below the lasing threshold.

We are now able to obtain an expression for the output ASE field. Solving the Heisenberg equations of motion for $b(\Delta,t)$, $s(\Delta,t)$ and $a(t)$ gives
\begin{equation}\label{b1}
\fl b(\Delta,t)=\exp\left[-\rmi\Delta(t-t_0)\right]b(\Delta,t_0)+\sqrt{\frac{\gbone}{2\pi}}\int_{t_0}^t \exp\left[-\rmi\Delta(t-\tau)\right]a(\tau)\,\rmd\tau
\end{equation}
\begin{equation}\label{s1}
\fl s(\Delta,t)=\exp\left[\rmi\Delta(t-t_0)\right]s(\Delta,t_0)-\sqrt{\frac{\gaone}{2\pi}}\int_{t_0}^t \exp\left[\rmi\Delta(t-\tau)\right]a^\dagger(\tau)\,\rmd\tau
\end{equation}
and
\begin{eqnarray}\label{at1}
\eqalign{\fl a(t)=-\int_{t_0}^t \exp\left[-\frac{\gbone-\gaone}{2}(t-\tau)\right]\left(\sqrt{\gbone}b_\text{in}(\tau)+\sqrt{\gaone}s_\text{in}^\dagger(\tau)\right)\,\rmd\tau\\
+\exp\left[-\frac{\gbone-\gaone}{2}(t-t_0)\right]a(t_0)}
\end{eqnarray}
where
\begin{equation}\label{bin}
\bin\equiv\frac{1}{\sqrt{2\pi}}\int_{-\infty}^\infty \exp\left[-\rmi\Delta(t-t_0)\right]b(\Delta,t_0)\,\rmd\Delta 
\end{equation}
is the radiation field that enters the cavity at time $t$ and
\begin{equation}\label{sin}
\sinput\equiv\frac{1}{\sqrt{2\pi}}\int_{-\infty}^\infty \exp\left[\rmi\Delta(t-t_0)\right]s(\Delta,t_0)\,\rmd\Delta 
\end{equation}
is the input atomic field. Time $t_0<0$ occurs at the beginning of region 1 in figure~\ref{RASEscheme}. Both input fields are vacuum fields with $\left<b^\dagger(\Delta,t_0) b(\Delta^\prime,t_0)\right>=\left<s^\dagger(\Delta,t_0) s(\Delta^\prime,t_0)\right>=0$. Note that we name the input atomic field from the point of view of the cavity mode; the input atomic field is the field that drives the cavity mode.
Equations~\eref{b1} and~\eref{s1} can be used to derive the relations~\cite{gardiner2004quantum}
\begin{equation}\label{binouta}
\base\equiv\bout=\bin+\sqrt{\gbone}a(t)
\end{equation}
\begin{equation}\label{sinouta}
\snout=\sinput-\sqrt{\gaone}a^\dagger(t)
\end{equation}
where
\begin{equation}
\bout\equiv\frac{1}{\sqrt{2\pi}}\int_{-\infty}^\infty \exp\left(-\rmi\Delta t\right)b(\Delta,0)\,\rmd\Delta 
\end{equation}
is the ASE field - the radiation field that exits the cavity at time $t$ - and
\begin{equation}\label{snout}
\snout\equiv\frac{1}{\sqrt{2\pi}}\int_{-\infty}^\infty \exp\left(\rmi\Delta t\right)s(\Delta,0)\,\rmd\Delta 
\end{equation}
is the output atomic field. The output atomic field is a collective de-excitation of atoms, or equivalently an excitation of the inverted harmonic oscillator field.

We will assume that the frequencies of interest of the ASE field is narrowband compared to $\gbone-\gaone$, the net loss rate of the cavity (see the implementation section for further discussion). We can then adiabatically eliminate the cavity mode, replacing $b_\text{in}(\tau)$ by $\bin$ and $s_\text{in}(\tau)$ by $\sinput$ in equation~\eref{at1}. Substituting the resulting expression for $a(t)$ into equations~\eref{binouta} and~\eref{sinouta} and letting $t_0\rightarrow -\infty$ gives the following input-output relations:

\begin{equation}\label{binout}
\base=-\frac{\gbone+\gaone}{\gbone-\gaone}\bin-\frac{2\sqrt{\gbone\gaone}}{\gbone-\gaone}\sind
\end{equation}
\begin{equation}\label{sinout}
\snout=\frac{\gbone+\gaone}{\gbone-\gaone}\sinput+\frac{2\sqrt{\gbone\gaone}}{\gbone-\gaone}\bind
\end{equation}
These relations allow us to determine the output ASE field and output atomic field from the known input radiation and atomic fields.

\section{The rephasing $\pi$-pulse}
The rephasing $\pi$-pulse (applied at $t=0$)  induces the following changes in our atoms~\cite{allen1975optical}:
\begin{equation}\label{szflip}
\sigma^k_z(0)\rightarrow -\sigma_z^k(\delta t)
\end{equation}
\begin{equation}\label{smflip}
\sigma_+^k(0)\rightarrow \sigma_-^k(\delta t)
\end{equation}
where $\delta t$ is the duration of the $\pi$-pulse and $\sigma_z^k(t)=2\sigma_+^k(t)\sigma_-^k(t)-I$, where $I$ is the identity operator. The $\pi$-pulse inverts the excitation of and rephases the atoms. Like in~\cite{ledingham2010nonclassical}, we model the rephasing pulse as an instantaneous $\pi$-pulse and so take $\delta t\rightarrow 0$. This is valid if $(\delta t)^{-1}$ is large compared to the bandwidth of the ASE and RASE fields that we are interested in.

\section{The RASE field}
After the rephasing $\pi$-pulse the atoms are predominantly in their ground state (equation~\eref{szflip}). But the atoms that fell to their ground state in region 1 will be in their excited state and these atoms produce the RASE field. Because the atoms are only weakly excited we approximate the atoms as a field of ordinary harmonic oscillators by setting $\sigma_-(\Delta,t)\rightarrow d(\Delta,t)$. Here $d(\Delta,t)$ are destruction operators satisfying $[d(\Delta,t),d^\dagger(\Delta^\prime,t)]=\delta(\Delta-\Delta^\prime)$.

The interaction picture Hamiltonian for our system takes the form (making the rotating wave approximation and setting $\hbar\equiv 1$)
\begin{eqnarray}
\eqalign{\fl H&=\int_{-\infty}^\infty \Delta b^\dagger(\Delta,t)b(\Delta,t)\,\rmd\Delta+\rmi\sqrt{\frac{\gbtwo}{2\pi}}\int_{-\infty}^\infty\left(b^\dagger(\Delta,t)a(t)-b(\Delta,t)a^\dagger(t)\right)\,\rmd\Delta\\
\fl &\phantom{equals}+\int_{-\infty}^\infty \Delta d^\dagger(\Delta,t)d(\Delta,t)\,\rmd\Delta+\rmi\sqrt{\frac{\gatwo}{2\pi}}\int_{-\infty}^\infty\left(d^\dagger(\Delta,t)a(t)-d(\Delta,t)a^\dagger(t)\right)\,\rmd\Delta}
\end{eqnarray}
The loss rate of the bare cavity is $\gbone$ and $\gaone$ is the rate the atoms would absorb photons from the cavity if the cavity mirrors were perfect.
We have allowed for the atom-cavity and radiation-cavity couplings to differ from the couplings for times $t<0$. This allows for the case where either the $Q$-factor of the cavity changes or the oscillator strength of the atomic transition used for the ASE and RASE are different~\cite{beavan2011photon}.

We again make the first Markov approximation and assume that the frequencies of interest of the RASE field is narrowband compared to $\gatwo+\gbtwo$, the total loss rate of the cavity, allowing us to again adiabatically eliminate the cavity mode. Carrying out analogous calculations to those from the previous section we obtain the following input-output relations:

\begin{equation}\label{binout2}
\brase\equiv\bout=-\frac{\gbtwo-\gatwo}{\gbtwo+\gatwo}\bin-\frac{2\sqrt{\gbtwo\gatwo}}{\gbtwo+\gatwo}\din
\end{equation}
\begin{equation}\label{dinout}
\dout=\frac{\gbtwo-\gatwo}{\gbtwo+\gatwo}\din-\frac{2\sqrt{\gbtwo\gatwo}}{\gbtwo+\gatwo}\bin
\end{equation}
where
\begin{equation}
\brase\equiv\frac{1}{\sqrt{2\pi}}\int_{-\infty}^\infty \exp\left[-\rmi\Delta(t-t_1)\right]b(\Delta,t_1)\,\rmd\Delta 
\end{equation}
is the output RASE field, where $t_1$ is any time far in the future,
\begin{equation}
\dout\equiv\frac{1}{\sqrt{2\pi}}\int_{-\infty}^\infty \exp\left[-\rmi\Delta(t-t_1)\right]d(\Delta,t_1)\,\rmd\Delta 
\end{equation}
is the output atomic field,
\begin{equation}\label{din}
\din\equiv\frac{1}{\sqrt{2\pi}}\int_{-\infty}^\infty \exp\left(-\rmi\Delta t\right)d(\Delta,0)\,\rmd\Delta 
\end{equation}
is the input atomic field and $\bin$ is defined by equation~\eref{bin}.

Equation~\eref{binout2} gives the RASE field as a function of the known input radiation field and the currently unknown input atomic field. The input atomic field for region 2 can be related to the output atomic field from region 1 using equation~\eref{smflip} with $\sigma_+^k(0)\rightarrow s(\Delta,0)$ and $\sigma_-^k(\delta t)\rightarrow d(\Delta,\delta t)$. The definitions of $\din$ and $\snout$ (equations~\eref{din} and~\eref{snout}) then give that
\begin{equation}\label{dinsnout}
\din={s_\text{out}}(-t)
\end{equation}
in the limit $\delta t\rightarrow 0$. Equation~\eref{dinsnout} shows that the input atomic field that drives the cavity mode after the rephasing pulse is equal to the output atomic field that was driven by the cavity mode before the rephasing pulse. Entanglement between $\snout$ and $\base$ before the rephasing pulse is translated into entanglement between $\din$ and $b_\text{ASE}(-t)$ as a result of equation~\eref{dinsnout}. The field $\brase$ becomes entangled with $\din$ and is therefore also entangled with $b_\text{ASE}(-t)$.

\begin{figure}[h]
\begin{center}
\includegraphics{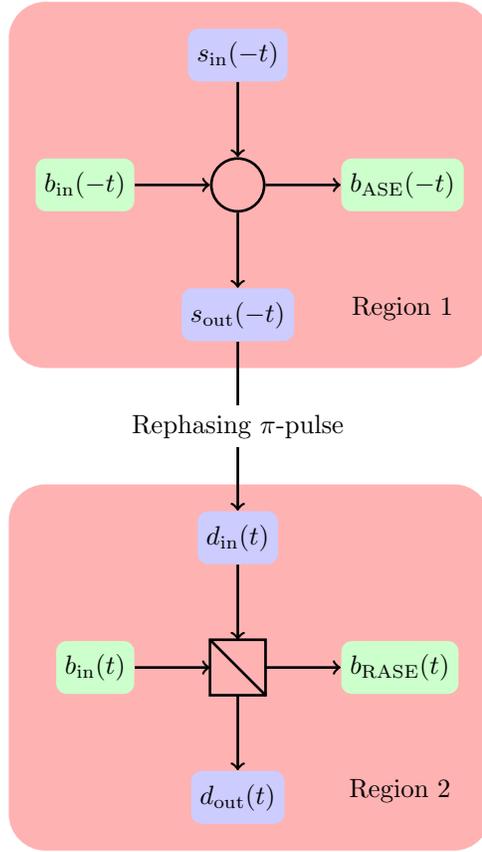}
\end{center}
\caption{\label{inputoutput}Graphical representation of the input-output relations~\eref{binout} and~\eref{brase}. In the time region 1 the input optical field $b_\text{in}(-t)$ and the input atomic field $s_\text{in}(-t)$ combine like in a non-degenerate parametric amplifier to produce the outputs $b_\text{ASE}(-t)$ and $s_\text{out}(-t)$. The rephasing pulse means that the input atomic field for region 2 ($\din$) matches the output atomic field from region 1 ($s_\text{out}(-t)$) . In region 2, the optical and atomic fields interact like on a beamsplitter. For an impedance matched cavity the reflectivity of this beamsplitter is 1. It is well known that in an impedance matched cavity full of atoms any input field gets totally absorbed by the atoms, and none of it escapes as light. What also happens in an impedance matched cavity is that the input atomic field is completely mapped onto the output optical field leading to 100\% recall efficiency. This means that if the cavity is impedance matched in region 2, the two output optical fields $b_\text{ASE}(-t)$ and $b_\text{RASE}(t)$ will be maximally entangled.}
\end{figure}

Using equations~\eref{dinsnout} and~\eref{sinout} in equation~\eref{binout2} gives
\begin{equation}\label{brase}
\fl \brase=-\frac{\gbtwo-\gatwo}{\gbtwo+\gatwo}\bin-\frac{2\sqrt{\gbtwo\gatwo}}{\gbtwo+\gatwo}\left(\frac{\gbone+\gaone}{\gbone-\gaone}s_\text{in}(-t)+\frac{2\sqrt{\gbone\gaone}}{\gbone-\gaone}b_\text{in}^\dagger(-t)\right)
\end{equation}
and this along with equation~\eref{binout} give the RASE and ASE fields as functions of the input atomic field before the rephasing pulse and input radiation field. The input-output relations describing our system are shown graphically in figure~\ref{inputoutput}.

\section{Quantifying entanglement}
To concentrate on a single temporal mode for both the ASE and RASE fields we introduce mode operators  $A$ and $B$ defined by
\begin{equation}
A\equiv-\int_{-\infty}^0 g(t)\base\,\rmd t
\end{equation}
\begin{equation}
B\equiv\int_0^\infty g(t)^*\brase\,\rmd t
\end{equation}
where $g(t)$ is a temporal mode function satisfying $g(-t)=g(t)$ and $\int_0^\infty |g(t)|^2\,\rmd t=1$. The number of photons in the ASE and RASE modes are then given by $N_\text{ASE}=\left<A^\dagger A\right>$ and $N_\text{RASE}=\left<B^\dagger B\right>$

Equations~\eref{binout} and~\eref{brase} give that
\begin{equation}\label{A}
A=A_0 \cosh\chi+B_0^\dagger\sinh\chi
\end{equation}
\begin{equation}\label{B}
B=\sqrt{\epsilon}C_0-\sqrt{1-\epsilon}\left(B_0\cosh\chi+A_0^\dagger\sinh\chi\right)
\end{equation}
where $A_0\equiv \int_{-\infty}^0 g(t) \bin\,\rmd t$, $B_0\equiv \int_{-\infty}^0 g(t)^* s_\text{in}(t)\,\rmd t$, $C_0\equiv -\int_0^\infty g(t)^*\bin\,\rmd t$, $\cosh \chi\equiv (\gbone+\gaone)/(\gbone-\gaone)$ and $\sqrt{\epsilon}\equiv (\gbtwo-\gatwo)/(\gbtwo+\gatwo)$. We have that $[c_j,c_k^\dagger]=\delta_{jk}$ and $\left<c_j^\dagger c_k\right>=0$, with $c_j\in \left\{A_0,B_0,C_0\right\}$. Equations~\eref{A} and~\eref{B} take the form of the equations describing the output of a non-degenerate parametric amplifier and beam splitter combination~\cite{walls1995quantum}, see figure~\ref{inputoutput}.

Equations~\eref{A} and~\eref{B} give that $N_\text{ASE}=\sinh^2\chi$ and $N_\text{RASE}=(1-\epsilon)\sinh^2\chi$. The recall efficiency of our system is given by $N_\text{RASE}/N_\text{ASE}=1-\epsilon$. If the cavity is impedance matched in region 2  ($\gatwo=\gbtwo$) the recall efficiency is 100\%, resulting in perfect entanglement between the ASE and RASE fields.

\begin{figure}[h]
\begin{center}
\includegraphics[width=\textwidth]{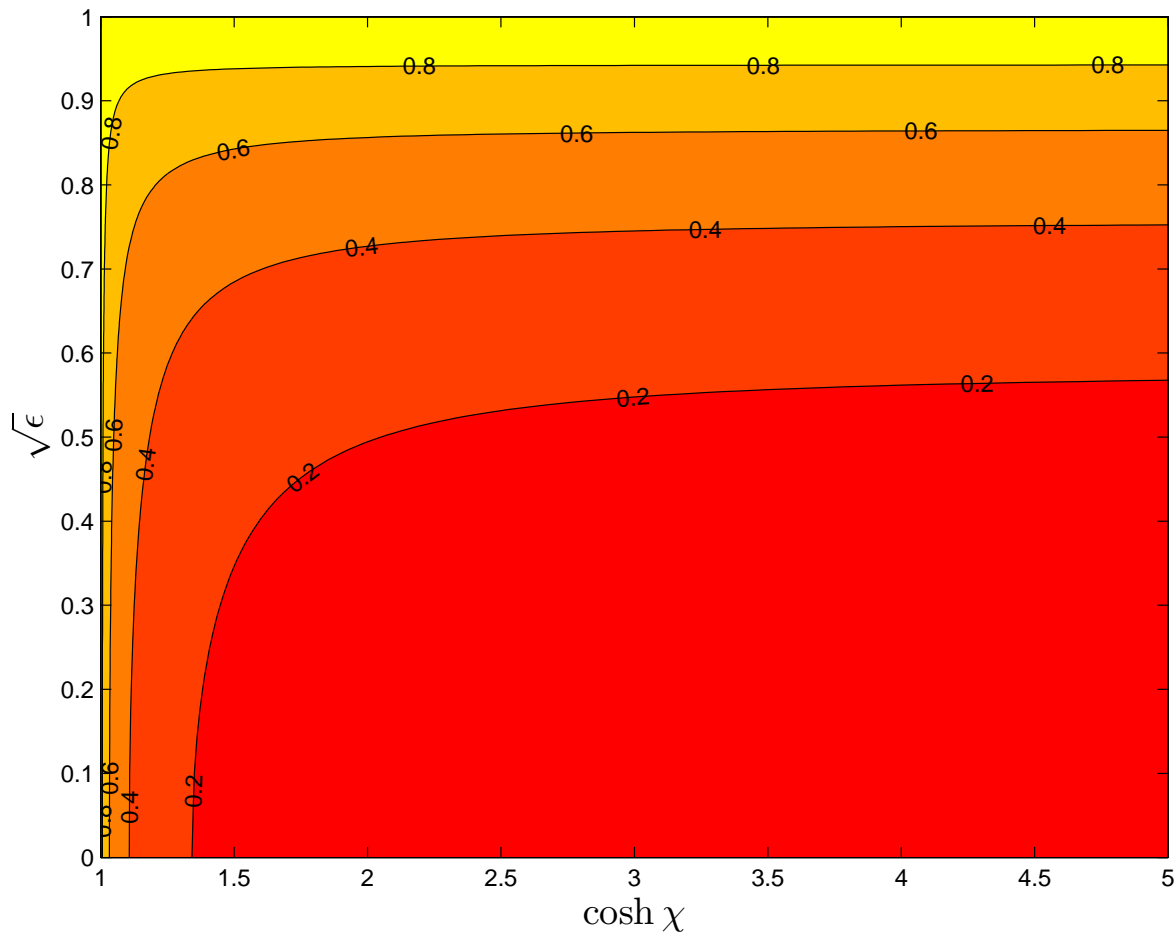}
\end{center}
\caption{\label{varUVplot}Contour plot of $\left<\Delta u^2\right>+\left<\Delta v^2\right>$, minimised with respect to $\theta$, versus $\sqrt{\epsilon}$ and $\cosh\chi$. The ASE and RASE fields are necessarily entangled in regions where $\left<\Delta u^2\right>+\left<\Delta v^2\right><1$. The ASE and RASE fields are entangled for all parameter values considered in this figure.}
\end{figure}

We will quantify the entanglement between the ASE and RASE fields using the criterion of Duan et al.~\cite{Duan2000}. For this criterion we introduce the operators
\begin{equation}
u\equiv\sqrt{\theta}\xa+\sqrt{1-\theta}\xb
\end{equation}
\begin{equation}
v\equiv \sqrt{\theta}\pa-\sqrt{1-\theta}\pb
\end{equation}
where $x_c\equiv (c+c^\dagger)/\sqrt{2}$ and $p_c\equiv -\rmi(c-c^\dagger)/\sqrt{2}$ are amplitude and phase quadrature fields satisfying $[x_c,p_{c^\prime}]=\rmi\delta_{c c^\prime}$, with $c,c^\prime\in\left\{\text{A},\text{B}\right\}$, and $\theta$ can be any real number in the interval $(0,1)$. Then a sufficient condition for entanglement between the ASE and RASE photons is that $\left<\Delta u^2\right>+\left<\Delta v^2\right>< 1$ \footnote{The condition given in~\cite{Duan2000} is that the ASE and RASE photons are entangled if $\left<\Delta u^2\right>+\left<\Delta v^2\right><\lambda^2+1/\lambda^2$, where $u\equiv |\lambda| \xa+(1/\lambda)\xb$ and $v=|\lambda|\pa-(1/\lambda)\pb$ for some real, non-zero $\lambda$. Setting $\theta\equiv\lambda^2(\lambda^2+1/\lambda^2)^{-1}$ gives the condition used in this paper.}. Figure \ref{varUVplot} shows a contour plot of $\left<\Delta u^2\right>+\left<\Delta v^2\right>$, minimised with respect to $\theta$, versus $\sqrt{\epsilon}$ and $\cosh\chi$. The ASE and RASE fields are entangled for all parameter values considered in this figure.

It can be shown that for our system
\begin{equation}
\fl \left<\Delta u^2\right>+\left<\Delta v^2\right>=1+2\sinh^2\chi-2\epsilon(1-\theta)\sinh^2\chi-4\sqrt{\theta-\theta^2}\sqrt{1-\epsilon}\cosh\chi\sinh\chi
\end{equation}
Setting $\theta=(1-\epsilon)/(2-\epsilon)$ gives $\left<\Delta u^2\right>+\left<\Delta v^2\right><1$ for all valid values of $\chi$ and $\epsilon$. Therefore the ASE and RASE fields are entangled for all valid parameter values.

\section{Implementation}
To implement our proposal, it is necessary to use two-level atoms with coherence times longer than the combined duration of the ASE and RASE fields. This could be achieved using the long coherence time of rare-earth doped solids~\cite{thiel2011rare} or by using Raman transitions in atomic gases~\cite{hetet2008photon}.

Operation with the cavity impedance matched in region 2 requires either some way of changing the atom oscillator strength, such as the four-level scheme of~\cite{beavan2011photon}, or a $Q$-switched cavity because, in order not to cross the lasing threshold, $\gbone>\gaone$ is required in region 1. In the case of the four-level scheme, the atom-cavity coupling in regions 1 and 2 can be made to differ by a factor $\gtrsim 10$ using appropriate rare-earth materials~\cite{nilsson2004hole, guillot2009hyperfine}. In the case of $Q$-switching, it is not satisfactory to simply insert a variable attenuator, as attenuation introduces an unwanted additional output port that reduces the photon recall efficiency. Instead, various methods of active $Q$-switching can be used that do not introduce unwanted loss: using AOM or EOM components allows for variable control of the output field~\cite{yariv1989quantum, svelto1998principles}; the output of a WGM resonator coupled to an optical fiber or prism can be varied by changing the distance between the WGM resonator and the fiber or prism~\cite{gorodetsky1999optical, gorodetsky1994high}; or a thin Fabry-P{\'e}rot resonator with variable mirror separation can be used as a mirror with a variable reflectivity~\cite{strain1994experimental}.

The cavity finesses required for implementation are only moderate, being no larger than that required to achieve impedance matching. This does not pose a great challenge, since the atomic absorption can be made much larger than the limit of current cavity decay rates.

The bandwidth of the output light is determined by either the cavity linewidth or the width of the inhomogeneous broadening. We will only be interested in a band of this light that is narrow compared to $\min\left(\gbone-\gaone, \gbtwo+\gatwo\right)$, since adiabatic elimination of the cavity dynamics is only valid for such a range. Filtering this narrowband signal from the full bandwidth of the output light can be achieved in a number of ways. The easiest to implement is homodyne or heterodyne detection~\cite{yuen1983noise}. In this way the filtering problem is reduced to filtering electrical signals. This method is suitable for quantum repeaters using a continuous variable basis~\cite{hage2008preparation,braunstein2005quantum}. For quantum repeaters using a discrete Fock basis, filtering using standard techniques such as Fabry-P{\'e}rot etalons is a possibility, and when using rare-earth ions one can also use the array of narrow band filtering techniques recently developed based on spectral holeburning~\cite{li2008pulsed,mcauslan2012using}.

It has been assumed that the rephasing $\pi$-pulse is a perfect $\pi$-pulse. The feasibility of applying a $\pi$-pulse to the whole ensemble is greatly helped by the area theorem~\cite{mccall1969self}, which states that a $\pi$-pulse will remain a $\pi$-pulse as it travels through resonant media. This has strong analogues in a cavity~\cite{chaneliere2013strong}. Of course, in practice the $\pi$-pulse won't be perfect. An imperfect $\pi$ pulse can be considered as a combination of a perfect $\pi$ pulse and a small perturbing pulse. The small perturbing pulse adds unwanted excitation to the atomic field, which adds noise to the RASE field. However, as discussed in~\cite{ledingham2010nonclassical}, this noise disappears shortly after the $\pi$ pulse, since the unwanted excitation of the atomic field will be temporally brief and will rapidly dephase and no longer interact with the cavity mode. The variation in driving strength of the atoms due to variations in the cavity mode field intensity could be removed using hole burning techniques~\cite{pryde2000solid,longdell2004experimental}.

\section{Conclusion}
In conclusion, we have analysed rephased amplified spontaneous emission with the atoms placed in an optical cavity. The cavity can alleviate the problem of low recall efficiency, particularly if the cavity is impedance matched when the entangled light is being recalled from the atoms, in which case the recall will theoretically be perfect. Achieving the impedance matched condition during recall requires either a $Q$-switched cavity or some way of switching the atoms' oscillator strengths, such as in the four-level scheme of Beavan et al.~\cite{beavan2011photon}, since the ASE field has to be produced below the lasing threshold. We have also shown that entanglement exists between the ASE and RASE fields for all valid parameter values. Theoretically, our system has the potential to achieve time-separated entanglement with perfect recall efficiency, which is indispensable in producing an effective quantum repeater.

\section*{Acknowledgements}
The authors would like to acknowledge financial support from the Marsden Fund of the Royal Society of New Zealand.

\section*{References}
\providecommand{\newblock}{}

\end{document}